\documentclass[intlimits,twoside,a4paper]{article}

\usepackage[cp1251]{inputenc}

\usepackage[eqsecnum]{cmpj3}

\usepackage{bm}

\issue{2021}{24}{2}{23602}
\doinumber{10.5488/CMP.24.23602}

\title[Band inversion in Na$_{2}$AgX (X= As, Sb and Bi)]%
{A first-principles investigation of band inversion in topologically nontrivial Na$_{2}$AgX (X= As, Sb and Bi) full Heusler compounds}%

\author[A. Boughena \textsl{et al.}]{A. Boughena\refaddr{label1}, S. Benalia\refaddr{label1, label2}, O. Cheref\refaddr{label1}, N. Bettahar\refaddr{label1}, D. Rached\refaddr{label1}}
\addresses{
\addr{label1} Laboratoire des Mat\'{e}riaux Magn\'{e}tiques, Facult\'{e} des Sciences Exactes, Universit\'{e} Djillali Liab\`{e}s de Sidi Bel-Abb\`{e}s, Sidi Bel-Abb\`{e}s (22000), Alg\'{e}rie
\addr{label2} D\'{e}partement des Sciences de la Mati\`{e}re, Facult\'{e} des Sciences et de la Technologie, Universit\'{e} Ahmed Ben Yahia El-Wancharisi Tissemsilt, Tissemsilt (38000), Algeria
}
\date{Received April 05, 2021, in final form May 13, 2021}

\begin{document}

\maketitle

\begin{abstract}
Topological nontrivial nature are the latest phases to be discovered in condensed matter physics with insulating bulk band gaps and topologically protected metallic surface states; they are one of the current hot topics because of their unique properties and potential applications. In this paper, we have highlighted a first-principles study of the structural stability and electronic behavior of the Na${}_{2}$AgX (X= As, Sb and Bi) full Heusler compounds, using the Full-Potential Linear Muffin-Tin Orbital (FP-LMTO) method. We have originated that the Hg${}_{2}$CuTi structure is appropriate in all studied materials. The negative values of the calculated formation energies mean that these compounds are energetically stable. The band structure is studied for the two cases relating the existence and the absence of spin-orbital couplings, where all materials are shown to be topologically non-trivial compounds. Spin orbital couplings were noticed to have no significant effect on the electronic properties such as the topological order.

\end{abstract}

\section{Introduction}

Topological nontrivial compounds have earned a great deal of attention~\cite{1,2} as they have become the most important topic in condensed matter physics~\cite{2,3} due to their unusual and exotic electronic properties. It is a new quantum state of matter that has surface states without gaps within the bulk energy gap~\cite{4,5}. There were produced several applications of spin-based electronic devices (spintronics) to topological insulators~\cite{6}. In general, topological insulators have very important physical advantages~\cite{7}.

Recently, the search for topological nontrivial compounds has extended to ternary Heusler and chalcopyrite compounds~\cite{8,9}. Moreover, a series of theoretical and experimental efforts have been dedicated to predict new Heusler topological insulators ({TIs}), in view of their great potential for spintronics and quantum computing applications~\cite{10,23}.  Furthermore, it should be emphasized that additionally to the fact that these materials provide a new interesting context  which makes it possible to identify and understand the physical consequences of topological properties of momentum-space bands or real-space texture. They also provide a  tempting prospect of adapting the discovered fundamental advances into  important new applications~\cite{24}. Interest in them is increasing continuously because of their multifarious properties for spintronic  applications~\cite{13},  optoelectronic~\cite{14}, superconductivity~\cite{15}, shape memory~\cite{16}, giant magneto resistance spin valve (GMR)~\cite{17}, thermoelectric applications~\cite{18}, and spin injection to semiconductors~\cite{19,21}.

 Motivated by their potential applications in spintronics and quantum computing, the search in three-dimensional topological nontrivial compounds based on Heusler compounds has attracted considerable theoretical and experimental interest~\cite{8,25,26,27,28,29}. In this review, we focus on the topological nontrivial compounds-based full Heusler alloys. A wide range of possible composition variations in the full Heusler crystal structure is expected to allow for highly tunable and versatile electronic properties, such as spin-structured topological surface states and topological superconductivity. Motivated by this review, and, with the aim of filling this knowledge gap, we have conducted an investigation on the band topological ordering of some newly Na${}_{2}$AgX (X= As, Sb and Bi) full Heusler alloys, where we  study their structural and electronic properties with and without including the spin-orbit coupling effect in our calculations. These calculations  help us, mainly, to classify the topological states of these materials and identify the exact class they belong to.

 The rest of the paper is arranged as follows: section 2 includes computational details and the calculation method, section 3 is devoted to the results regarding the structural and electronic properties. Discussion, a brief conclusion and an outlook is drawn in section 4.
\section{Calculation methodology}

Calculations for our work were performed with and without spin-orbit coupling  (SOC) effects by adopting  the full-potential linear muffin tin orbital (FP-LMTO) method~\cite{31,32} based on  the density functional theory (DFT)~\cite{33,34} by performing the local-density approximation (LDA)~\cite{35,36}.  In order to achieve the energy eigenvalues convergence, the charge density and potential inside the MTSs are represented by spherical harmonics up $l_{\text{max}}= 6$.  The self-consistent calculations are considered to be converged when the total energy of the system is stable at $10^{-6}$~Ry. The $k$ integration over the Brillouin zone is performed using the tetrahedron method~\cite{37}. To avoid the overlap of atomic spheres, the MTSs radii for each atomic position are taken differently for each composition. We point out that performing the full-potential calculations ensures  the calculations being not completely independent of the choice of the spheres radii.

\section{Results and discussion}

\begin{figure}[!t]
\centerline{\includegraphics[width=4.58in, height=2.85in]{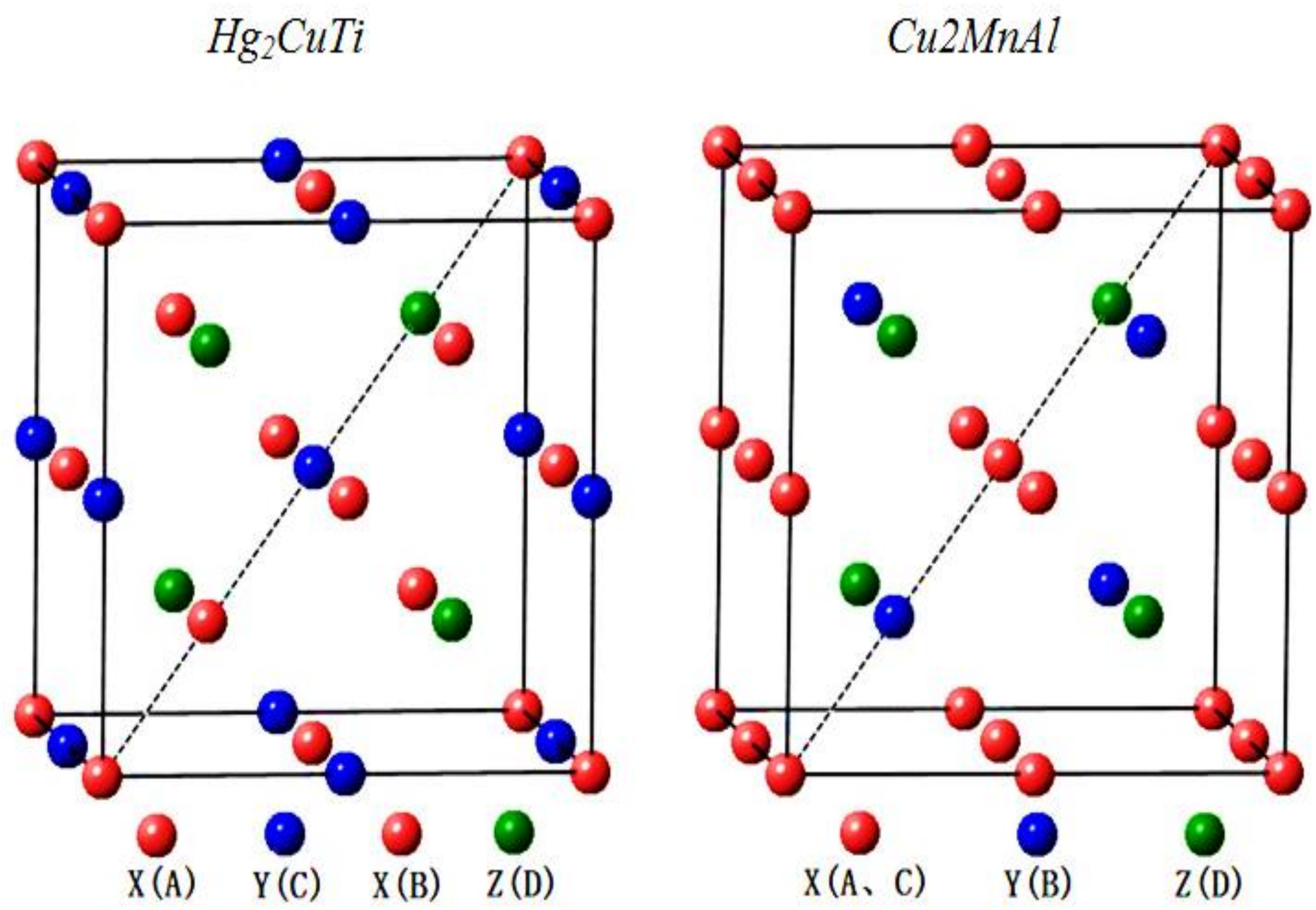}}
\caption{(Colour online) Crystal structures of full Heusler compound, Hg$_{2}$CuTi: in which X atoms are placed in A~(0,~0,~0) and B~($1/4$,~$1/4$,~$1/4$), Y and Z atoms are placed in C~($1/2$,~$1/2$,~$1/2$) and D~($3/4$,~$3/4$,~$3/4$) Wyckoff positions, respectively, and Cu$_{2}$MnAl: in which X atoms are occupied A~(0,~0,~0) and C~($1/2$,~$1/2$,~$1/2$) sites, Y and Z atoms are occupied B~($1/4$,~$1/4$,~$1/4$) and D ($3/4$,~$3/4$,~$3/4$) sites, respectively.} \label{figure 1}
\end{figure}

In the case of ordered cubic Heusler materials, two types of prototypical structures are recognized. The so-called ``regular'' type (prototype Cu${}_{2}$MnAl, L21), which crystallizes in space group (Fm-3m~N$^\circ$225), with three unavoidable Wyckoff positions [4(a), 4(b), 4(c) and 4(d)] incorporating four atoms per unit cell. Wyckoff positions 4(a) and 4(b) are occupied by two Na atoms with coordinates (0,~0,~0) and ($1/2$,~$1/2$,~$1/2$) respectively, the positions 4(c) with coordinates ($1/4$,~$1/4$,~$1/4$) and 4(d) with coordinates ($3/4$,~$3/4$,~$3/4$) are filled with Ag and X (X= As, Bi and Sb) atoms, respectively. The coordination of the closest neighborhood is shown in figure~\ref{figure 1}. The second prototype often appearing in the context of Heusler compounds is the Hg${}_{2}$CuTi type structure (sometimes called ``inverted Heusler'', figure~\ref{figure 1}), which can be derived from the ``regular'' Heusler type by exchanging the atoms on the 4(c) position with the element occupying the 4(b) position. This reduces the symmetry of the cell, leading to space group F-43m N$^\circ$216 (type XA) with four unequal positions in the unit cell. In this case, these are occupied by Na at 4(a) and 4(c) with coordinates (0,~0,~0) and ($1/4$,~$1/4$,~$1/4$), respectively, Ag at 4(b) with coordinates ($1/2$,~$1/2$,~$1/2$), and X (X= As, Bi and Sb) at 4(d) with coordinates ($3/4$,~$3/4$,~$3/4$). We have determined the different values of the structural parameters at steady state such as the lattice constants ($a_{0}$) and the corresponding total energy ($E_{0}$), bulk moduli ($B$) and its pressure derivatives ($B'$) for the two possible structural configurations of type  L21 and X-type using the fitting Murunaghan's equation of state~\cite{38}. We have also calculated the formation energy ($E_{f}$)  for these Heusler compounds at equilibrium state. All these results obtained are listed in table~\ref{Table 1}. To the best of our knowledge, there is no theoretical or experimental work exploring the lattice constants ($a_{0}$) of these compounds. The Na$_{2}$AgX (X= As, Sb and Bi) compounds are being predicted for the first time in this work.

On the other hand, we have found in all the studied materials that the minimum energy corresponds to the Hg$_{2}$CuTi structure. Therefore, the stable structure of all full Heusler compound is the X-type. Rather, it should be noted that the energy difference between the two structures (L21 and X-type) in their structural parameters corresponds to the variations in the position of the Na and Ag atoms. This means that these atoms play an important role in the stabilization of the X-type structure. Furthermore, it can also be noted that this energy difference slightly increases  with an increase of the volume of the lattice. Having this in mind, we note that the lattice parameter $a_{0}$ (table~\ref{Table 1}) increases when replacing As by Bi and Sb. This is due to the variation of the atomic number of the element X [Z(Bi)$>$Z(Sb)$>$Z(As)] because there is an obvious correlation between the lattice parameter and the atomic number of the X element. Moreover, the average formation energy of a material is the energy needed to separate its components into free atoms. It is a measure of the intensity of the force that binds to the set of atoms that correlate with the structural stability of the ground state. The formation energy per unit cell is given by:

\[{E}^{\rm{Na}_2 \rm{AgX}}_{f}={E}^{\rm{Na}_2\rm{AgX}}_{\rm{Tot}}-\left( 2E^{\rm{Na}}_{\rm{Bulk}}+E^{\rm{Ag}}_{\rm{Bulk}}+E^{\rm{X}}_{\rm{Bulk}}\right),\] 
where $E_{\rm{Tot}}$ [energy total of Na$_{2}$AgX (X= As, Sb and Bi)] is the total energy for the ternary Heusler compounds taken at equilibrium, $E^{\rm{Na}}_{\rm{Bulk}}$, $E^{\rm{Ag}}_{\rm{Bulk}}$, $E^{\rm{X}}_{\rm{Bulk}}$ are the total energies of the free atoms for Na, Ag, and X (X= As, Sb and Bi), respectively. Accordingly, the results (table~\ref{Table 1}) show that the calculated formation energies have negative values; this means that these compounds are energetically stable, and, therefore, we conclude that these compounds could be synthesized experimentally.

\begin{table}[htb]
\caption{Calculated lattice constants, bulk moduli, their pressure derivatives, total energies ($E_{0}$) and formation energies ($E_{f}$) of Na$_{2}$AgX (X= As, Sb and Bi) full Heusler compounds.}
\label{Table 1}
\vspace{3mm}
\begin{center}
\begin{tabular}{||c c c c c c c||} 
 \hline
 Compound & Structure & $a_{0}$ (a.u) & $B$ (GPa) & $B'$ & $E_{0}$ (Ryd) & $E_{f}$(eV) \\ [0.5ex] 
 \hline
 Na${}_{2}$AgAs & L2${}_{1}$ & 12.8098 & 44.8411 & 4.3819 & $-15787.71101$ & /\\ 
 
 Na${}_{2}$AgAs & X-type & 12.6148 & 56.1616 & 4.1427 & $-15787.77554$ & $-11.038$\\ 
 \hline
 Na${}_{2}$AgBi & L2${}_{1}$ & 13.5317 & 39.8424 & 4.4007 & $-54411.66405$ & /\\ 
 
 Na${}_{2}$AgBi & X-type & 13.4615 & 51.457 & 3.68489 & $-54411.71669$ & $-10.63$\\ 
 \hline
 Na${}_{2}$AgSb & L2${}_{1}$ & 13.2843 & 38.8132 & 4.7395 & $-24227.9377$ & /\\ 
 
 Na${}_{2}$AgSb & X-type & 13.0668 & 46.1642 & 4.8659 & $-24227.99506$ & $-8.523$\\ 
 \hline
\end{tabular}

\end{center}
\end{table}

Studying the band structure is a critical step in understanding the behavior of materials, as well as in identifying the class which these materials belong to~\cite{39}. It can also be used to identify the topological insulating nature of the studied materials by  noticing the existence of band inversions within the electronic structure~\cite{40, 41}. Spin-orbit couplings (SOC) that cause the band splitting are widely assumed to be responsible for the band inversion at $\Gamma$ point. Thus, driven by this assumption, the search for new topological insulating materials is mainly fixed on the compounds with strong SOC due to the contained heavy elements~\cite{42, 43}. However, the band inversion has been also shown to be induced by other factors such as lattice strain, and without any strong SOC. Moreover, and surprisingly, the band inversion in HgTe has been demonstrated to be non-dependent on the strength of the SOC~\cite{42}. This result is very indicative since it offers a wider plethora for the search of TIs, going beyond materials with strong SOC, as well as  it motivates the study of the effect of SOC on the topology of materials, with the aim of identifying whether they are responsible for band inversion or not. Motivated by this, we have investigated the band structure and topological character of the target full-Heusler alloys. 

\begin{figure}[!b]
\centerline{\includegraphics[width=4.58in, height=5.85in]{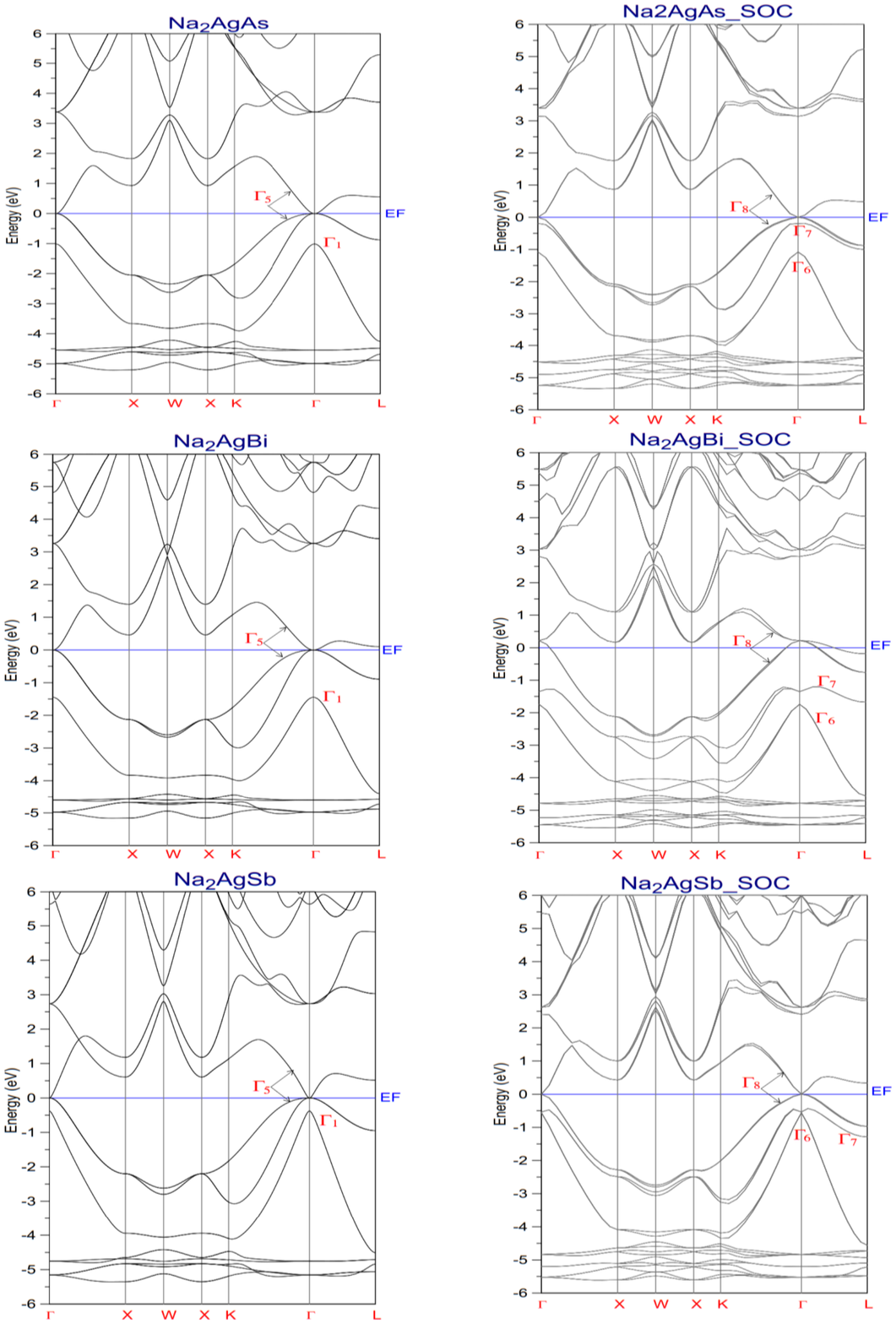}}
\caption{(Colour online) Band structure of Na${}_{2}$AgX (X= As, Sb and Bi) full Heusler compounds using LDA approach with and without SOC.} \label{figure 2}
\end{figure}

Figure~\ref{figure 2} shows that for the three compounds Na$_{2}$AgX (X=~As, Sb and Bi), the band structure is similar to the structure  HgTe,  thus showing a clear evidence of band inversion, where the $p$-like band ($\Gamma_{5}$) is above the $s$-like states band ($\Gamma_{1}$), indicating the non-trivial nature of the phase of these materials. Since the calculations are done without considering strong SOC, these results demonstrated that SOC are not a prime responsible agent in instigating the band inversion in the electronic structure of these compounds. By including the SOC effect into the calculations, we obtain the band structures shown in figure~\ref{figure 2}. The band inversion is shown to be preserved, without any clear change in the band inversion. The $\Gamma_{5}$ state is split into the quartet $\Gamma_{8}$ and the doublet $\Gamma_{7}$ states, while the $\Gamma_{1}$ state is changed to the doublet $\Gamma_{6}$ state. The characteristic inverted band order at the $\Gamma$ point is clearly observed. Specifically, the $s$-like $\Gamma_{6}$ state sits under the $p$-like $\Gamma_{8}$ state and the $E_{\rm{F}}$. Therefore, these alloys are topologically nontrivial compounds in their ground states with semimetallic behavior~\cite{44}.

\section{Conclusion}
In this work, we carried out a detailed study of the structural and electronic properties of the Na$_{2}$AgX (X= As, Sb and Bi) full Heusler compounds, using the FP-LMTO method.

\begin{enumerate}
\item  We have found in all the studied materials that the minimum energy corresponds to the Hg$_{2}$CuTi structure. Therefore, the stable structure of all full Heusler compound is the X-type.

\item  The calculated formation energies have negative values. This means that these compounds are energetically stable, and, therefore, we conclude that these compounds could be synthesized experimentally.

\item  The band structure is  similar to the HgTe structure,  thus showing a clear evidence of band inversion, where the $p$-like band ($\Gamma_{5}$) is above the $s$-like states band ($\Gamma_{1}$), indicating the non-trivial nature.

\item  By including the SOC effect into the calculations, the band inversion is shown to be preserved, without any clear change in the band inversion. The $\Gamma_{5}$ state is split into the quartet $\Gamma_{8}$ and the doublet $\Gamma_{7}$ states, while the $\Gamma_{1}$ state is changed to the doublet $\Gamma_{6}$ state.

\item  These results demonstrated that SOC are not a prime responsible agent in instigating the band inversion in the electronic structure of these compounds.
\end{enumerate}

\newpage
\ukrainianpart

\title[Інверсія зон в Na$_{2}$AgX (X= As, Sb і Bi)]%
{Першопринципне дослідження інверсії зон в топологічно нетривіальних повних гейслерівських сполуках 
Na$_{2}$AgX\\(X= As, Sb і Bi)}
\author{А. Бугена \refaddr{label1}, С. Беналія\refaddr{label1, label2}, 
О. Шереф \refaddr{label1}, Н. Беттагар\refaddr{label1}, Д. Рашид\refaddr{label1}}
\addresses{
\addr{label1} Лабораторія магнітних матеріалів, факультет точних наук, 
університет Джиллалі Ліабес, 22000 Сіді Бель-Аббес, Алжир
\addr{label2} Факультет науки та технології, університет Ахмед бен Ях'я Ель-Ванхарізі, 38000 Тіссемсілт, Алжир
}

\makeukrtitle

\begin{abstract}
	Топологічно нетривіальні фази у природі є одними з небагатьох ще недосліджених у фізиці конденсованого стану  
	з діелектричними об'ємними забороненими зонами і топологічно захищеними металічними поверхневими станами; їх вивчення є досить актуальним з огляду на
	унікальні властивості цих об'єктів та їх потенційні застосування. У даній статті висвітлено результати першопринципних досліджень структурної стійкості та 
	електронної поведінки гейслерівських сполук Na$_{2}$AgX (X= As, Sb та Bi) з використанням повнопотенціального лінійного методу орбіталей іонного остову (ПЛМ-ОІО). 
	Виявлено, що  структура типу Hg$_{2}$CuTi є прийнятною для усіх досліджених сполук. Від'ємні значення  розрахованих енергій утворення означають, шо ці сполуки є 
	енергетично стійкими. Зонну структуру досліджено для двох випадків --- присутніх та відсутніх спін-орбітальних взаємодій --- і показано, що усі досліджені матеріали
	є топологічно нетривіальними сполуками. Зауважено, що спін-орбітальні взаємодії не мають значного впливу на електронні властивості цих матеріалів, такі як 
	топологічний порядок.
	\keywords метод функціоналу густини, гейслерівські сполуки, спін-орбітальні взаємодії, топологічний порядок, 
	електронні властивості
\end{abstract}


\begin{thebibliography}{44}

\bibitem{1} Wang~X.~T., Dai~X.~F., Jia~H.~Y., Wang~L.~Y., Liu~G.~D., Liu~X.~F., Yuan W., Cui~Y.~T., Rare Met., 2021, \textbf{40}, 1219--1223, \doi{10.1007/s12598-014-0421-1}.

\bibitem{2} Jhon~Y.~I., Lee~J., Jhon~Y.~M., Lee~J.~H., IEEE J. Sel. Top. Quantum Electron., 2018, \textbf{24}, No. 5, 1--8,\\ \doi{10.1109/JSTQE.2018.2811903}.

\bibitem{3} Tokura~Y., Yasuda~K., Tsukazaki~A., Nat. Rev. Phys., 2019, \textbf{1}, 126--143, \doi{10.1038/s42254-018-0011-5}.

\bibitem{4}  M\"{u}chler~L., Casper~F., Yan~B., Chadov~S., Felser~C., Phys. Status Solidi RRL, 2013, \textbf{7}, No. 1--2, 91--100,\\ \doi{10.1002/pssr.201206411}. 

\bibitem{5}  Xu~Y., Gan~Z., Zhang~S.~C., Phys. Rev. Lett., 2014, \textbf{112}, No. 22, 226801, \doi{10.1103/PhysRevLett.112.226801}. 

\bibitem{6}  Wang Y., Gedik N., Phys. Status Solidi RRL, 2013, \textbf{7}, No. 1--2, 64--71, \doi{10.1002/pssr.201206458}.  

\bibitem{7}  Zhang~H., Zhang~S.~C., Phys. Status Solidi RRL, 2013, \textbf{7}, No. 1--2, 72--81, \doi{10.1002/pssr.201206414}.

\bibitem{8}  Chadov S., Qi X.-L., Kubler J., Fecher J. H., Felser C., Zhang S.-C., Nat. Mater., 2010, \textbf{9}, 541--545,\\ \doi{10.1038/nmat2770}.
\bibitem{9}  Feng W., Xiao D., Ding J., Yao Y., Phys. Rev. Lett., 2011, \textbf{106}, No. 1--7, 016402,\\ \doi{10.1103/PhysRevLett.106.016402}.

\bibitem{10} Heusler~F., Verh. Dtsch. Phys. Ges., 1903, \textbf{5}, 219.

\bibitem{23} Moore~J.~E., Nat. Phys., 2009, \textbf{5}, 378--380, \doi{10.1038/nphys1294}.

\bibitem{24} \v{S}mejkal~L., Mokrousov~Y., Yan~B., MacDonald~A.~H., Nat. Phys., 2018, \textbf{14}, 242--251, \\\doi{10.1038/s41567-018-0064-5}.



\bibitem{13} Felser~C., Fecher~G.~H., Balke~B., Angew. Chem. Int. Ed., 2007, \textbf{46}, No. 5, 668--699, \doi{10.1002/anie.200601815}.

\bibitem{14} Kieven~D., Klenk~R., Naghavi~S., Felser~C., Gruhn~T., Phys. Rev. B, 2010, \textbf{81}, No. 5, 075208,\\ \doi{10.1103/PhysRevB.81.075208}.

\bibitem{15} Winterlik~J., Fecher~G.~H., Felser~C., Solid State Commun., 2008, \textbf{145}, No. 9--10, 475--478,\\ \doi{10.1016/j.ssc.2007.12.020}.

\bibitem{16} Blum C. F. G., Ouardi~S., Fecher~G.~H., Balke~B., Kozina~X., Stryganyuk~G., Ueda~S., Kobayashi~K., Felser~C., Wurmehl~S., B\"{u}chner~B., Appl. Phys. Lett., 2011, \textbf{98}, 252501, \doi{10.1063/1.3600663}.

\bibitem{17} Dieny~B., Speriosu~V.~S., Parkin~S.~S.~P., Gurney~B.~A., Wilhoit~D.~R., Mauri~D., Phys. Rev. B, 1991,  \textbf{43}, 1297, \doi{10.1103/PhysRevB.43.1297}.

\bibitem{18} Heusler~Fr., Z. Anorg. Allg. Chem., 1927, \textbf{161}, 159--160.

\bibitem{19} Ohno~Y., Young~D.~K., Beshoten~B., Matsukura~F., Ohno~H., Awschalom~D.~D., Nature, 1999, \textbf{402}, 790--792, \doi{10.1038/45509}.

\bibitem{21} Fiederling~R., Keim~M., Reuscher~G., Ossau~W., Schmidt~G., Waag~A., Molenkamp~L.~W., Nature, 1999, \textbf{402}, 787--190, \doi{10.1038/45502}.

\bibitem{25} Yan~B., Liu~C.-X., Zhang~H.-J., Yam~C.-Y., Qi~X.-L., Frauenheim~T., Zhang~S.-C., Europhys. Lett., 2010, \textbf{90}, No.~3, 37002, \doi{10.1209/0295-5075/90/37002}. 

\bibitem{26} Lin~H., Markiewicz~R.~S., Wray~L.~A., Fu~L., Hasan~M.~Z., Bansil~A., Phys. Rev. Lett., 2010, \textbf{105}, No.~3--16, 036404, \doi{10.1103/PhysRevLett.105.036404}.

\bibitem{27} Sato~T., Segawa~K., Guo~H., Sugawara~K., Souma~S., Takahashi~T., Ando~Y., Phys. Rev. Lett., 2010, \textbf{105}, No.~3--16, 136802, \doi{10.1103/PhysRevLett.105.136802}.

\bibitem{28} Chen~Y. L., Liu~Z. K., Analytis~J.~G., Chu~J.-H., Zhang~H. J.,  Yan H., Mo~S.-K., Moore~R.~G., Lu~D. H., Fisher~I. R., Zhang~S. C., Hussain~Z., Shen~Z.-X., Phys. Rev. Lett., 2010,  \textbf{105}, 266401, \doi{10.1103/PhysRevLett.105.266401}.


\bibitem{29} Lin~H., Wray~L.~A., Xia~Y., Xu~S., Jia~S., Cava~R.~J., Bansil~A., Hasan~M.~Z., Nat. Mater., 2010, \textbf{9}, 546--549,\\ \doi{10.1038/nmat2771}.





\bibitem{31} Savrasov~S.~Yu., Savrasov~D.~Yu., Phys. Rev. B, 1992, \textbf{46}, 12181, \doi{10.1103/PhysRevB.46.12181}.

\bibitem{32} Savrasov~S.~Y., Phys. Rev. B, 1996, \textbf{54}, 16470, \doi{10.1103/PhysRevB.54.16470}.
\bibitem{33} Hohenberg~P., Kohn~W., Phys. Rev., 1964, \textbf{136}, 864, \doi{10.1103/PhysRev.136.B864}.

\bibitem{34} Kohn~W., Sham~L.~J., Phys. Rev., 1965, \textbf{140}, A1133, \doi{10.1103/PhysRev.140.A1133}.

\bibitem{35} Perdew~J.~P., Wang~Y., Phys. Rev. B, 1992, \textbf{45}, 13244, \doi{10.1103/PhysRevB.45.13244}.

\bibitem{36}  Perdew~J.~P., Wang~Y., Phys. Rev. B, 1992, \textbf{46}, 12947, \doi{10.1103/PhysRevB.46.12947}.

\bibitem{37} Bl\"ochl~P.~E., Jepsen~O., Andersen~O.~K., Phys. Rev. B, 1994, \textbf{49}, 16223, \doi{10.1103/PhysRevB.49.16223}.
\bibitem{38} Murnaghan F., Proc. Natl. Acad. Sci. U.S.A., 1944, \textbf{30}, 244--247, \doi{10.1073/pnas.30.9.244}.


\bibitem{39} Singleton J., Band Theory and Electronic Properties of Solids, Oxford Univ. Press, New York, 2001.

\bibitem{40} Lin H., Wray~A., Xia~Y., Xu~S., Jia~S., Cava~R.~J., Bansil~A., Hasan~M.~Z., Nat. Mater., 2010, \textbf{9}, 546--549,\\ \doi{10.1038/nmat2771}.

\bibitem{41} Singh A.~K., Ramarao~S., Peter~S.~C., APL Mater., 2020, \textbf{8}, 060903, \doi{10.1063/5.0006118}.

\bibitem{42} Zhu Z., Cheng~Y., Schwingenschl\"{o}gl~U., Phys. Rev. B, 2012, \textbf{85}, 235401, \doi{10.1103/PhysRevB.85.235401}.

\bibitem{43} Qi~X.-L., Zhang~S.-C., Rev. Mod. Phys., 2011, \textbf{83}, 1057, \doi{10.1103/RevModPhys.83.1057}.

\bibitem{44} Zhang~T., Jiang~Y., Song~Z., Huang~H., He~Y., Fang~Z., Weng~H., Fang~C., Nature, 2019, \textbf{566}, 475--479,\\ \doi{10.1038/s41586-019-0944-6}. 



\bibitem{20} Schmidt~G., Ferrand~D., Molenkamp~L.~W., Filip~A.~T., van~Wees~B.~J., Phys. Rev. B, 2000, \textbf{62}, \\R4790, \doi{10.1103/PhysRevB.62.R4790}.

\bibitem{22} Fu~L., Kane~C.~L., Mele~E.~J., Phys. Rev. Lett., 2007, \textbf{98}, 106803, \doi{10.1103/PhysRevLett.98.106803}.

\bibitem{11} Chadov~S., Qi~X., K\"{u}bler~J., Fecher~G.~H., Felser~C.~S., Zhang~S.~C., Nat. Mater., 2010, \textbf{9}, 541--545,\\ \doi{10.1038/nmat2770}.


\bibitem{12} Lin~H., Wray~L.~A., Xia~Y., Xu~S., Jia~S., Cava~R.~J., Nat. Mater., 2010, \textbf{9}, 546--549, \doi{10.1038/nmat2771}.








\end{thebibliography}
\end{document}